\documentclass[]{iopart}
\usepackage{iopams}
\usepackage{setstack}
\usepackage[]{graphicx}
\usepackage{color}
\usepackage[latin1]{inputenc}

\begin{document}

\title[]{On the average of inconsistent data}
\author{G Mana and M Predescu}
\address{INRIM -- Istituto Nazionale di Ricerca Metrologica, str.\ delle Cacce 91, 10135 Torino, Italy}

\begin{abstract}
When data do not conform to the hypothesis of a known sampling-variance, the fitting of a constant to the set of measured values is a long debated problem. Given the data, the fitting would require to find which measurand value is most probable. A fitting procedure is here reviewed which assigns probabilities to the possible measurand values, on the assumption that the uncertainty associated with each datum is the lower bound to the standard deviation. This procedure is applied to derive an estimate of the Planck constant.
\end{abstract}

\submitto{Metrologia}

% 06.20.Dk Measurement and error theory  
% 02.50.Cw Probability theory
% 07.05.Kf Data analysis: algorithms and implementation; data management
% 06.20.Jr Determination of fundamental constants 

%\baselineskip 8mm

\section{Introduction}
Given a set of measured values of a constant and the associated uncertainties, the Gauss-Markov theorem states that the unbiased minimum-variance estimator of the measurand is the weighted mean \cite{Luemberger}. The uncertainty of the mean, which is smaller than the smallest uncertainty, does not depend on data spread. This is a consequence of the assumption that the variance of the sampling distribution of each datum is exactly know. In practice, this hypothesis is often false and the inconsistency of the data -- quantified, for example, by the $\chi^2$ or the Birge-ratio values -- suggests that the uncertainties associated with the data are merely lower bounds to the standard deviations of their sampling distributions. In this case, the Gauss-Markov theorem is of no help and the choice of an optimal measurand value is long debated issue. The consequences of the assumption that the associated uncertainties are point estimates of the standard deviations are illustrated by Dose \cite{Dose}, who considers the estimate of the Newtonian constant of gravitation.

Decision theory and probability calculus help to deal with the discrepancy between the quoted uncertainties and the data scatter. Given the measurement results and the lower bounds of the standard deviations, the first step is to find the probabilities of the possible measurand values. As described by Sivia \cite{Sivia}, who solved the related problem of dealing with outliers, probabilities are assigned by application of the product rule and marginalization. Next, given the loss due to a wrong decision, the optimal choice of the measurand value minimizes the expected loss over the assigned probabilities. Since foundations are in the probability theory, this method makes it no longer necessary to exclude the data disagreeing with the majority, as well as to scale the uncertainties to make the data consistent. After reviewing the Sivia's analysis, the procedure is here applied to estimate the value of the Planck constant from a set of inconsistent measurement results.

\section{Problem statement}
Let us consider a set of $N$ measured values $x_i$ of a measurand $h$, which have been independently sampled from Gaussian distributions having unknown variance $\sigma_i^2 \ge u_i^2$, where $u_i$ is the uncertainty associated with the $x_i$ datum. The problem is to find optimal estimates of the measurand value and of the confidence intervals. Clearly, we cannot rely on the weighted mean, because the actual variances of the sampling distributions are not known. On the other hand, we cannot go back to the arithmetic mean, because it leaves out significant information delivered by the associated uncertainties.

\section{Problem solution}\label{problem}
Given the data and their associated uncertainty, the solution is to assign probabilities to the $h$ values and to use probabilities to minimize any given loss function \cite{Jaynes,McKay,Gregory}.

Firstly, we make probability assignments to the possible $\sigma_i$ values, for which only the $u_i$ lower bounds are known. In the absence of any additional information, we assume that the probability distribution of $\sigma_i$ is independent of the measurement unit, that is, $aP(a\sigma)=P(\sigma)$, where $P(\sigma)$ is the sought distribution of the $\sigma$ values \cite{Sivia}. This implies
\begin{equation}\label{prior}
 P(\sigma;u) = \theta(\sigma-u)/\sigma ,
\end{equation}
where the Heaviside function $\theta(\sigma-u)$ is 1 if $\sigma \ge u$ and zero if $\sigma < u$ and the $i$ subscript has been dropped. The distribution (\ref{prior}) is improper, but, provided the final measurand distribution is integrable, this is not a serious problem. Anyhow, we can always set a finite upper bound and delay the calculation of the upper bound limit to the infinity until last.

Next, having stated $P(\sigma;u)$, the sampling distribution of each $x$ datum can be marginalized to eliminate the unknown variance. Hence,
\begin{equation}\label{sampling}
 Q(x;h,u) = \int_u^\infty N(x;h,\sigma^2)P(\sigma;u)\, \rmd \sigma = \frac{{\rm erf}\big[(x - h)/\sqrt{2u^2}\big]}{2(x - h)} ,
\end{equation}
where $N(x;h,\sigma^2)$ is a normal distribution of the $x$ values having $h$ mean and $\sigma^2$ variance and erf is the error function.

Eventually, we make pre-data probability assignments to the $h$ values. In the absence of any additional information, we assume that they are independent of the unit-scale origin, which implies a uniform distribution.

By application of the product rule of probability, the only post-data probability distribution of the measurand values, logically consistent with the pre-data assignments and sampling distributions, is
\begin{equation}\label{measurand}
 \Theta(h) = \frac{1}{Z} \prod_{i=1}^N Q(x_i;h,u_i) ,
\end{equation}
where the $Q$ function is given in (\ref{sampling}), $1/Z$ is a normalization factor, $x_i$ and $u_i$ are the $i$-th datum and its uncertainty.

The post-data distribution (\ref{measurand}) is the central to our analysis. Given any loss $L(h_0-h)$ due to a wrong estimate $h_0$, the optimal choice of the $h$ value minimizes
\begin{equation}\label{cost}
 \langle L(h_0-h) \rangle = \int_{-\infty}^{+\infty} L(h_0-h) \Theta(h)\, \rmd h .
\end{equation}
With a quadratic loss, the optimal estimate is the mean; with a linear loss, it is the median. With a loss independent of the error, it is the mode. Confidence intervals are easily calculable from (\ref{measurand}).

\begin{table}[b]
\caption{Values of the Planck constant. The input data are from \cite{CODATA} and the references therein, with the exception of the last two, that are from \cite{Rainville:2005}. The values calculated from the quotient of $h$ and the neutron mass, the mass of a particle or of an atom, and the neutron binding energy depend on the same measured value of $N_A$ \cite{Andreas}.}
\begin{center}
\begin{tabular}{ll}
\hline\hline
method &$h$ / $10^{-34}$ Js  \\
\hline\hline
 quotient of $h$ and the electron charge \\
 $h/(2e)$  (NMI 89)                 &$6.6260684(36)  $ \\
 $h/(2e)$  (PTB 91)                 &$6.6260670(42)  $ \\
\hline
 watt-balance experiments \\
 $h/m_{\mathfrak{K}}$ (NPL 1990)    &$6.6260682(13)  $ \\
 $h/m_{\mathfrak{K}}$ (NIST 1998 )  &$6.62606891(58) $ \\
 $h/m_{\mathfrak{K}}$ (NIST 2007)   &$6.62606901(34) $ \\
 $h/m_{\mathfrak{K}}$ (METAS 2011)  &$6.6260691(20)  $ \\
 $h/m_{\mathfrak{K}}$ (NPL 2010)    &$6.6260712(13)  $ \\
\hline
 quotient of $h$ and the neutron mass \\
 $N_A h/M({\rm n})$                 &$6.62606887(52) $ \\
\hline
 quotient of $h$ and the mass of a particle or of an atom \\
 $N_A h/M(e)$                       &$6.62607003(20) $ \\
 $N_A h/M({\rm Cs})$                &$6.62607000(22) $ \\
 $N_A h/M({\rm Rb})$                &$6.62607011(22) $ \\
 weighted mean                      &$6.62607005(20) $ \\
\hline
 quotient of $h$ and the neutron binding energy \\
 $h/\Delta m(^{29}{\rm Si})$  &$6.6260764(53) $  \\
 $h/\Delta m(^{33}{\rm S})$   &$6.6260686(34) $  \\
 weighted mean                &$6.6260709(29) $  \\ \hline\hline
\end{tabular} \label{planck-values-table} \end{center} \end{table}

\section{Planck constant estimate}
As an application example, let us consider the choice of the Planck constant value on the basis of the measurment results listed in table \ref{planck-values-table}. We neither presume to account for the interlinks between the data nor we presume to compete with the adjustments carried out by the committee on data for science and technology \cite{CODATA}.

\begin{figure}
\centering
\includegraphics[width=75mm]{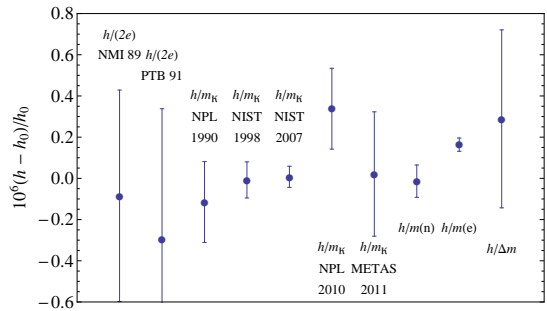}
\caption{Values of the Planck constant given in table \ref{planck-values-table}. The reference is the CODATA 2006 value, $h_0= 6.62606896 \times 10^{-34}$ Js \cite{CODATA}.}\label{planck-values-plot}
\end{figure}

The Planck constant links energy and momentum to the frequency and wavelength of the, classical or relativistic, wave-function. Therefore, $h$ determinations match energy (momentum) and frequency (wavelength) measurements. According to whether an electrical or mechanical system is considered, the measurement result is a value of the $h/e$ or $h/m$ ratio, where $e$ and $m$ are the electron charge and mass.

The electric determinations are based on the measurement of $h/(2e)$ by the tunneling of Cooper's pairs in a Josephson junction. The watt-balance experiments accede directly to the $h/m_\mathfrak{K}$ ratio, where $m_\mathfrak{K}$ is the mass of the international kilogram prototype. The measurement of the Avogadro constant $N_A$ opened the way towards comparing the watt-balance measurements of $h$ with the $h/m$ quotients, where $m$ is mass of a particle or of an atom. The $h/m(n)$ ratio was determined by time-of-flight measurements of monochromatic neutrons; optical spectroscopy and atom interferometry were instrumental to determine $h/m(e)$, $h/m({\rm Cs})$, and $h/m({\rm Rb})$. Nuclear spectroscopy of $^{28}$Si and $^{32}$S, after the capture of a thermal neutron, allowed the $N_A h$ product to be determined by comparing the mass defect with the frequencies of the $\gamma$ photons emitted in the cascades from the capture state to the ground state.

\begin{table}[b]
\caption{Mode, mean, and median values of the Planck constant, given the data in table \ref{planck-values-table}. The weighted-mean of the data -- whose associated uncertainties have been expanded to make the Birge ratio unit -- is also given, with the relevant standard deviation. The uncertainties associated to the mode and mean are the standard deviations of the post-data probability distribution; the sub- and super-script of the median indicate the extrema of the first and third quartiles.}
\begin{center}
\begin{tabular}{ll}
\hline\hline
mode        &6.62606923(60) \\
mean        &6.62606937(60) \\
median      &$6.62606931_{-38}^{+40}$ \\
\hline
weighted mean &6.62606960(20) \\
\hline\hline
\end{tabular} \label{h_values} \end{center} \end{table}

\begin{figure}
\centering
\includegraphics[width=75mm]{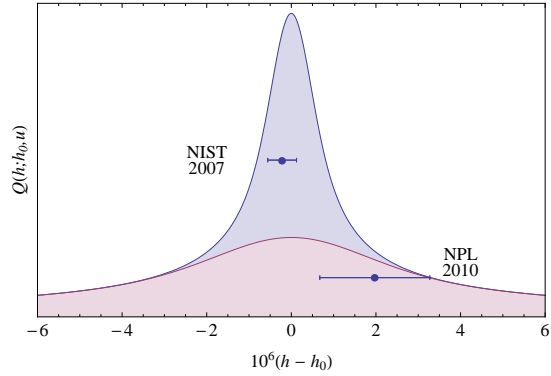}
\caption{Marginal distributions of the NIST 2007 and NPL 2010 measured values given lower bounds of the standard deviations equal to the associated uncertainties. The different areas under the curves are due to the improper character of these distributions. The measured values are also shown with their associated uncertainty. The reference value is the mode $h_0= 6.62606923 \times 10^{-34}$ Js of the post-data distribution shown in Fig.\ \ref{post}.}\label{fig-sampling}
\end{figure}

Figure \ref{planck-values-plot} shows the input data used in the analysis. The values calculated from the measurements of $h/m(e)$, $h/m({\rm Cs})$, and $h/m({\rm Rb})$ have been averaged, because their uncertainty is mainly affected by the $N_A$ determination and, therefore, they are highly correlated. The values obtained from nuclear spectroscopy have been averaged because they were obtained by repetitions of the same experiment.

Though none of them is totally out of scale, the data are inconsistent. The weighted mean is $6.62606960 \times 10^{-34}$ J/s with $\chi^2 = 14$ for $\nu=N-1=9$ degrees of freedom and Birge ratio $R_B=\sqrt{\chi^2/\nu} = 1.26$. In order to make the data consistent, it is usually assumed that the quoted uncertainties are wrong by a common scale factor. By multiplying the data uncertainties by the Birge ratio, the $\chi^2$ of the weighted mean is 9 and the Birge ratio is 1; this removes the discrepancy between data spread and uncertainties. The weighted mean and the final associated uncertainty are given in table \ref{h_values}.

\begin{figure}[b]
\centering
\includegraphics[width=70mm]{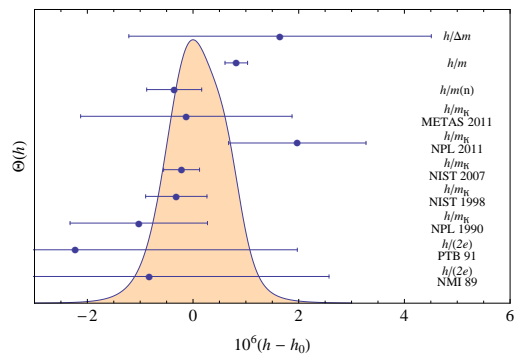}
\caption{Post-data probability density of the values of the Planck constant calculated on the basis of the measurement results given in table \ref{planck-values-table}. The input data are shown with their associated uncertainty. The reference value is the distribution mode, $h_0= 6.62606923 \times 10^{-34}$ Js.}\label{post}
\end{figure}

When assuming that the associated uncertainties are the lower bounds to unknown standard-deviations of the data, the probability calculus, as shown in section \ref{problem},  allows a probability to be assigned to each possible $h$ value. According to (\ref{measurand}), the post-data distribution of the $h$ values is
\begin{equation}\label{d-post}
 \Theta(h) = \frac{1}{Z} \prod_{i=1}^N \frac{{\rm erf}\big[(x_i - h)/\sqrt{2u_i^2}\big]}{2(x_i - h)} .
\end{equation}

By using the $h_0= 6.62606923 \times 10^{-34}$ Js mode of the post-data distribution (the actual $h$ value is unknown), the marginal sampling distributions, $Q(x;h_0,u_{\rm NIST07})$ and $Q(x;h_0,u_{\rm NPL10})$, of the NIST 2007 and NPL 2010 data are shown in Fig. \ref{fig-sampling}. The post-data distribution is shown in Fig.\ \ref{post}; the most probable $h$ value and its mean and median are given in table \ref{h_values}. It is to be noted that all these estimates are smaller than the weighted mean and that the standard deviation of the post-data distribution is significantly greater than that of the weighted mean.

\section{Conclusions}
When data do not conform to the hypothesis of a known variance, the probability calculus offers a solution to the problem of fitting a constant to the set of measured values. By assigning probabilities to the possible measurand values, measurand values and confidence intervals can be optimally chosen. This approach is more rigorous  than the search of estimators having optimal distributions for which no guideline exists. Assigning probabilities helps in the choice of a measurand value also when a traveling standard is used to compare the measurement capabilities of different laboratories. The post data distribution of the measurand values depends on the hypotheses made for the probability density of the unknown standard-deviation of data. The probability calculus allows different hypotheses to be compared; this will be the matter of future investigations.

\ack
This research received funding from the European Community's Seventh Framework Programme, ERA-NET Plus, under the iMERA-Plus Project - Grant Agreement No. 217257.


\section*{References}

\begin{thebibliography}{99}
\bibitem{Luemberger}
 Luemberger D G 1969 Optimization by Vector Space Methods (New York: John Wiley \& Sons, Inc.)
\bibitem{Dose}
 Dose V 2007 Bayesian estimate of the Newtonian constant of gravitation {\it Meas.\ Sci.\ Technol.} {\bf 18} 176-82
\bibitem{Sivia}
 D Sivia and J Skilling 2006 Data Analysis: A Bayesian Tutorial (Oxford: Oxford University Press)
\bibitem{Jaynes}
 Jaynes E T 2003 Probability theory: The logic of science (Cambridge: Cambridge University Press)
\bibitem{McKay}
 Mc Kay D JC 2003 Information Theory, Inference, and Learning Algorithms (Cambridge: Cambridge University Press)
\bibitem{Gregory}
 Gregory P C 2005 Bayesian Logical Data Analysis for the Physical Sciences (Cambridge: Cambridge University Press)
\bibitem{CODATA}
 Mohr P J, Taylor B N, and Newell D B 2008 CODATA recommended values of the fundamental physical constants: 2006 {\it Rev. Mod. Phys.} {\bf 80} 633-730
\bibitem{Rainville:2005}
 Rainville S {\it et al.} 2005 A direct test of $E=mc^2$ {\it Nature} {\bf 438} 1096-7
\bibitem{Andreas}
 Andreas B, Azuma Y, Bartl G, Becker P, Bettin H, Borys M, Busch I, Fuchs P, Fujii K, Fujimoto H, Kessler E, Krumrey M, Kuetgens U, Kuramoto N, Mana G, Massa E, Mizushima S, Nicolaus A, Picard A, Pramann A, Rienitz O, Schiel D, Valkiers S, Waseda A, and Zakel S 2011 Counting the atoms in a $^{28}$Si crystal for a new kilogram definition {\it Metrologia} {\bf 48} S1-S13
\end{thebibliography}
\end{document}